\documentclass[preprint,showpacs,11pt]{revtex4}

\usepackage{graphicx}
\usepackage{dcolumn}
\usepackage{bm}

\begin{document}


\title{On non-relativistic $Q\bar Q$ potential via Wilson loop in Galilean space-time}

\author{Haryanto M. Siahaan}%
 \email{haryanto.siahaan@gmail.com, haryanto.siahaan@home.unpar.ac.id}
\affiliation{
\small{Theoretical and Computational Physics Group, Department of Physics, Faculty
of Information Technology and Sciences, Parahyangan Catholic University,
Bandung 40141 - INDONESIA}}


\begin{abstract}
We calculate the static Wilson loop from string/gauge correspondence to obtain the $Q\bar Q$ potential in non-relativistic quantum field theory, i.e. CFT with Galilean symmetry. We analyze the convexity conditions \cite{bachas} for $Q\bar Q$ potential in this theory, and obtain restrictions for the acceptable dynamical exponent $z$.
\end{abstract}

\pacs{11.25.Tq}
\keywords{Wilson loop, Galilean symmetry, $Q\bar Q$ pair, potential, holography}
\maketitle

\section{Introduction}
It has been shown by Maldacena that large N superconformal gauge theories have a dual description in terms of string theory in AdS space \cite{Maldacena:1997re}. This proposal was realized by Maldacena to compute the energy between quark $(Q)$ and anti-quark $(\bar Q)$ pairs \cite{Maldawilson}. His method was to calculate expectation values of an operator similar to the Wilson loop in the large N limit of field theories. Maldacena's idea was improved later by Rey, Theisen, and Yee \cite{Rey:1998ik}. It turns Wilson loop into a physical gauge invariant property that can be read from the string picture. The $Q\bar Q$ energy in the large N superconformal ${\cal N} = 4$ Yang-Mills theory can be obtained from the Wilson loop of the corresponding string in AdS space. It is proposed that quark and anti-quark pairs live on the boundary, connected by a U-shaped string in the bulk. In the discussion on this spacetime, the energy has a non-confining Coulomb-like behavior, as expected for a conformal field theory. Later this approach was applied to many other spaces and models, as summarized in ref. \cite{sonwilson}.
\par Recently, gravity duals for a certain Galilean-invariant conformal field theory has attracted some attention in theoretical high energy physics community \cite{Hartnoll:2009sz,Son:2005rv,Nishida:2007pj,Son:2008ye,Balasubramanian:2008dm}.  A special case when we take the dynamical exponent $z = 2$ of this theory (whose isometry is the Schrodinger group $Sch(d-1)$) is considered to be the basis in constructing duality between gravity and unitary Fermi gas. However, our interest in this paper is the theory with an arbitrary dynamical exponent $z$, i.e. Galilean invariant CFT. In this general scheme, one can discuss the non-relativistic version of the AdS/CFT dictionary, i.e. the operator-state correspondence between the particle on the boundary and the string in the bulk. Scaling transformation in this non-relativistic conformal symmetry can be written as \cite{Son:2008ye,Balasubramanian:2008dm,Akhavan:2008ep}
\begin{eqnarray}
	x^i  \to \lambda x^i, t \to \lambda ^z t .\label{eq11}
\end{eqnarray}
\par The asymptotic metric in this case can be written as
\begin{eqnarray}
	ds^2  = \frac{{R^2 }}{{r^2 }}\left( { - \frac{{dt^2 }}{{r^{2\left( {z - 1} \right)} }} + dtd\xi  + \left( {dx^i } \right)^2 + dr^2 } \right) + ds_{X_5}^2 \label{eq12}
\end{eqnarray}
where $R$ is the characteristic radius of space-time, $\xi$ is a compact light-like coordinate, $x^i$ for $i = 1$,...,$d$ together with $t$ are the space-time coordinates on the boundary where (\ref{eq12}) is mapped at $r=0$, and finally $ds_{X_5}^2$ is the metric of a suitable internal manifold geometry which allows (\ref{eq12}) to be a solution of the supergravity equations of motion. The extra dimension $\xi$ is usually associated with quantum numbers interpreted as the particle number. However, the relation between translation in $\xi$ and its interpretation as particle number operator is still an unclear topic \cite{Kluson:2010prd,Balabsnumber2010}. Thus we just set this time-like extra dimension $\xi$ to be constant.
\par The holographic Wilson loop in non-relativistic CFT had been studied by Kluson in ref. \cite{Kluson:2010prd}. He assumed general time dependence of $\xi$ and also the moving $Q\bar Q$ pair cases in the context of non-relativistic quantum field theory. His study was devoted to the space-time with Galilean symmetry \footnote{From now on this will be abbreviated as Galilean space-time.}. Nevertheless, he still does not include analysis of convexity conditions (\ref{eq31}) and (\ref{eq32}) yet. 
One needs to verify these conditions in $Q\bar Q$ potential discussions to make sure that the corresponding potential function $V\left( L \right)$ is a monotone non-decreasing and convex function of the separation $L$. The goal of this paper is to verify these conditions for $Q\bar Q$ potential, which is obtained by calculating the Wilson loop in the string picture in Galilean space-time. Furthermore, we would like to see the restrictions which may appear for acceptable dynamical exponent $z$.

\par This paper is organized as follows. In section 2, we will perform calculations to acquire the $Q\bar Q$ potential energy in Galilean space-time. In section 3, we will derive some conditions for acceptable $z$ due to convexity inequality. Finally in section 4, there is a summary of our findings.

\section{$Q\bar Q$ potential in non-relativistic CFT with Galilean symmetry}

\begin{figure}[t]
        \centering 
	\includegraphics[width=8cm]{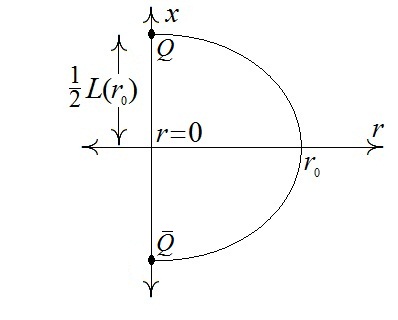}
	\caption{$Q\bar Q$ pair on the boundary as each ends of string.} 
	\label{qq2}
\end{figure}

We will start with the Nambu-Goto action 
\begin{eqnarray}
	S =  - \frac{1}{{2\pi \alpha '}}\int {d\tau d\sigma \sqrt { - \det G_{MN} \partial _\alpha  x^M \partial _\beta  x^N } } \label{eq21}
\end{eqnarray}
for metric (\ref{eq12}) where $
x^M  = \left( {t,r,\xi ,x^i } \right)$, $G_{MN}$ is space-time metric in (\ref{eq12}), and impose  suitable ansatzs in describing static strings, i.e. $t = x^0 = \tau $, $r = r\left( \sigma  \right)$, $x = x\left( \sigma  \right)$, and $\xi  = {\rm{constant}}$. Kluson in ref. \cite{Kluson:2010prd} has considered a more general case for an extra time-like dimension $\xi$ as a $\tau$-dependent variable, but we can simply set $\xi$  to be constant (for example as disucussed in ref. \cite{Akhavan:2008ep}) since the $Q\bar Q$ potential would depend on their separation distance \footnote{A distance between $Q$ and $\bar Q$ in our $3+1$ dimensional world, i.e. on the boundary of the Galilean bulk, see Fig.\ref{qq2}.} only. The corresponding action can be written as
\begin{eqnarray}
	S =  - \frac{T}{{2\pi \alpha '}}\int {d\sigma \sqrt {f^2 \left( r \right)\left( {\left( {r'} \right)^2  + \left( {x'} \right)^2 } \right)} } \label{eq22}
\end{eqnarray}
for $f\left( r \right) = R^2 r^{-(z+1)}$ and we have used $\left( {} \right)' \equiv \partial _\sigma  \left( {} \right)$. Variable $T$ in (\ref{eq22}) is the loop period and can be written this way due to the time translation invariance of action (\ref{eq21}) for metric (\ref{eq12}). We have followed a standard prescription that has been used in some literature, for example refs. \cite{sonwilson,nunez,filho,caceres,kinar,arias}, in obtaining the action (\ref{eq22}) as well as the corresponding $Q\bar Q$ potential as a function of $Q\bar Q$ pair's distance. Though the metric (2.1) is not diagonal, but action (\ref{eq22}) leads us to a problem of Wilson loop computation which can be started by finding a geodesic in the effective 2-dimensional geometry \cite{arias}
\begin{eqnarray}
	\left( {ds_{eff} } \right)^2  = f^2 \left( r \right)\left( {dx^2  + dr^2 } \right).\label{eq23}
\end{eqnarray}
\par The equation of motion (geodesic line) from (\ref{eq22}) is
\begin{eqnarray}
	\frac{{dx}}{{dr}} =  \pm \frac{{f\left( {r_0 } \right)}}{{\sqrt {f^2 \left( r \right) - f^2 \left( {r_0 } \right)} }}.\label{eq24}
\end{eqnarray}
$r_0$ is the maximum position of the U-shaped string with respect to the $r$-coordinate (bulk radius, see Fig. \ref{qq2}). From (\ref{eq24}) one can obtain the separation distance of quark and anti-quark on the boundary, by integrating the geodesic with respect to $r$. Since the boundary is at $r = 0$, then the separation as the function of $r_0$ can be obtained by the following integration
\begin{eqnarray}
	L\left( {r_0 } \right) = 2\int\limits_0^{r_0 } {\frac{{f\left( {r_0 } \right)}}{{\sqrt {f^2 \left( r \right) - f^2 \left( {r_0 } \right)} }}dr}. \label{eq25}
\end{eqnarray}
Related to the expression for the $Q\bar Q$ separation above, one may provide such an illustration as depicted in Fig. \ref{qq2}.

\par Inserting $f\left( r \right) = R^2 r^{-(z+1)}$ to (\ref{eq25}) and using the beta function in our computation give the following exact result
\begin{eqnarray}
	L\left( {r_0 ,z} \right) = 2\int\limits_0^{r_0 } {\frac{{r^{z + 1} }}{{\sqrt {r_0^{2z + 2}  - r^{2z + 2} } }}}  = \frac{{2r_0 \sqrt \pi  \Gamma \left( {\frac{{z + 2}}{{2z + 2}}} \right)}}{{\Gamma \left( {\frac{1}{{2z + 2}}} \right)}}.\label{eq26}
\end{eqnarray}
Then we follow a general prescription in refs. \cite{sonwilson,filho,kinar,arias} to compute the energy between quark and anti-quark. We have a general form of total $Q\bar Q$ energy as
\begin{eqnarray}
	E\left( {r_0 } \right) = \frac{1}{{\pi \alpha '}}\int\limits_0^{r_0 } {\frac{{f^2 \left( r \right)}}{{\sqrt {f^2 \left( r \right) - f^2 \left( {r_0 } \right)} }}dr}  - 2m_Q  \label{eq27}
\end{eqnarray}
where $m_Q$ is considered as the energy of non interacting quark \cite{nunez,filho,kinar,arias}. Thus the $Q\bar Q$ potential can be written as
	\[V_{Q\bar Q} \left( {r_0 } \right) = E\left( {r_0 } \right) - 2m_Q
\]
\begin{eqnarray}
	  = \frac{1}{{\pi \alpha '}}\int\limits_0^{r_0 } {\frac{{f^2 \left( r \right)}}{{\sqrt {f^2 \left( r \right) - f^2 \left( {r_0 } \right)} }}dr} \label{eq28}
\end{eqnarray}
which can also be computed by the use of beta function. The potential is
\begin{eqnarray}
	V_{Q\bar Q} \left( {r_0 ,z} \right) = 2R^2 r_0^{z + 1} \int\limits_0^{r_0 } {\frac{{dr}}{{r^{z + 1} \sqrt {r_0^{2z + 2}  - r^{2z + 2} } }}}  = \frac{{2R^2 \sqrt \pi  }}{{r_0 ^z \left( {2z + 2} \right)}}\frac{{\Gamma \left( {\frac{{ - z}}{{2z + 2}}} \right)}}{{\Gamma \left( {\frac{1}{{2z + 2}}} \right)}}.\label{eq29}
\end{eqnarray}
In the next section we will see the compatibility of the potential (\ref{eq29}) with convexity conditions.

\section{Convexity conditions and string embeddings}
There are some conditions that should be satisfied by any potential which describes interaction between quark and anti-quark whose name 'convexity' conditions \cite{bachas,arias}
\begin{eqnarray}
	\frac{{dV}}{{dL}} > 0 \label{eq31}
\end{eqnarray}
and
\begin{eqnarray}
\frac{{d^2 V}}{{dL^2 }} \le 0.\label{eq32}
\end{eqnarray}
Condition (\ref{eq31}) means quark and anti-quark are attractive everywhere, and (\ref{eq32}) tells us that the potential is a monotone non-increasing function of their separation. These conditions can be verified as follows 
\begin{eqnarray}
	\frac{{dV_{Q\bar Q} \left( {r_0 ,z} \right)}}{{dL\left( {r_0 ,z} \right)}} = \frac{{dV_{Q\bar Q} \left( {r_0 ,z} \right)}}{{dr_0 }}\frac{{dr_0 }}{{dL\left( {r_0 ,z} \right)}} = \frac{{ - zR^2 }}{{r_0 ^{z + 1} \left( {2z + 2} \right)}}\frac{{\Gamma \left( {\frac{{ - z}}{{2z + 2}}} \right)}}{{\Gamma \left( {\frac{{z + 2}}{{2z + 2}}} \right)}} > 0\label{eq33}
\end{eqnarray}
and
	\[\frac{{d^2 V_{Q\bar Q} \left( {r_0 ,z} \right)}}{{dL\left( {r_0 ,z} \right)^2 }} = \frac{{d\left( {\frac{{dV_{Q\bar Q} \left( {r_0 ,z} \right)}}{{dL\left( {r_0 ,z} \right)}}} \right)}}{{dr_0 }}\frac{{dr_0 }}{{dL\left( {r_0 ,z} \right)}} 
\]
\begin{eqnarray}
	= \frac{{zR^2 }}{{4\sqrt \pi  r_0 ^{z + 2} }}\frac{{\Gamma \left( {\frac{1}{{2z + 2}}} \right)\Gamma \left( {\frac{{ - z}}{{2z + 2}}} \right)}}{{\left( {\Gamma \left( {\frac{{z + 2}}{{2z + 2}}} \right)} \right)^2 }}\le 0.\label{eq34}
\end{eqnarray}
The two last equations are inequalities for physically accepted $z$ based on convexity conditions for the $Q\bar{Q}$ pair.

In ref. \cite{yoshida-hartnoll}, the authors present simple embeddings of duals for nonrelativistic critical points, where the dynamical critical exponent can take many values $z \ne 2$ \footnote{I thank Koushik Balasubramanian to give me know this work.}. They find that $z = 1$ and $z \ge 3/2$ as the possible dynamical critical exponents that allow string embeddings in gauge/gravity dual picture. From their paper \cite{yoshida-hartnoll}, we could learn that our $f\left( r \right)$ would depend on the coordinates of the internal manifold $X_{5}$ \footnote{I thank reviewer for pointing this out to me.}. Hartnoll and Yoshida write the non-compact part of the metric which can accommodate a large number of values of $z$ by the following ansatz \footnote{We follow the form of metric by Balasubramanian and McGreevy \cite{Balasubramanian:2008dm}. $f\left( {X_5 } \right)$ in ref. \cite{yoshida-hartnoll} is $h^2 \left( {X_5 } \right)$ in this paper.}
\begin{eqnarray}
	ds^2  = \frac{{R^2 }}{{r^2 }}\left( { - \frac{{dt^2 }}{{h^2 \left( {X_5 } \right)r^{2\left( {z - 1} \right)} }} + dtd\xi  + \left( {dx^i } \right)^2  + dr^2 } \right)\label{eq35}
\end{eqnarray}
which modifies our previous $ f\left( r \right) $ from $R^2 r^{ - (z + 1)} $ to $ R^2 r^{ - (z + 1)} h\left( {X_5 } \right)^{ - 1} $. Nevertheless, the function $ h\left( {X_5 } \right) $ would not appear in (\ref{eq26}) and (\ref{eq29}). Thus our findings on the restrictions for $z$ can be applied to the work of Hartnoll and Yoshida in ref. \cite{yoshida-hartnoll}. One can verify that conditions (\ref{eq33}) and (\ref{eq34}) are fulfilled for $z = 1$, and also for $z \ge 3/2$. The negativity of $\Gamma \left( {{\textstyle{{ - z} \over {2z + 2}}}} \right)$ for $z \ge 1$ guarantees both (\ref{eq33}) and (\ref{eq34}) are satisfied.

\section{Summary}
We have calculated the potential between $Q$ and $\bar Q$ in the non-relativistic quantum field theory by using the Wilson loop analysis in the gauge/gravity correspondence in the Galilean bulk. Our findings are inequalities (\ref{eq33}) and (\ref{eq34}) for physically acceptable dynamical exponent $z$ from convexity conditions. Yoshida and Hartnoll \cite{yoshida-hartnoll} have found families of $z$ for string embeddings in Galilean space-time, i.e. $z=1$ and $z \ge 3/2$, which agree with inequalities (\ref{eq33}) and (\ref{eq34}) above.

\section*{Acknowledgments}

I would thank LPPM-UNPAR for supporting my research. I also thank to my colleagues in physics department of Parahyangan Catholic University for all their supports, and to Frank Landsman of PPB-UNPAR for correcting my manuscript.



\begin{thebibliography}{99}
\bibitem{Maldacena:1997re}
  J.~M.~Maldacena,
\emph{``The large N limit of
 superconformal field theories and supergravity,''}
  Adv.\ Theor.\ Math.\ Phys.\  {\bf 2} (1998) 231
  [Int.\ J.\ Theor.\ Phys.\  {\bf 38} (1999) 1113]
  [arXiv:hep-th/9711200].

\bibitem{Maldawilson}
  J.~M.~Maldacena,
\emph{``Wilson loops in large N field theories,''}
  Phys.\ Rev.\ Lett.\  {\bf 80}, 4859 (1998)
  [arXiv:hep-th/9803002].

\bibitem{Rey:1998ik}
  S.~J.~Rey and J.~T.~Yee,
\emph{``Macroscopic strings as heavy
 quarks in large N gauge theory and  anti-de
  Sitter supergravity,''}
  Eur.\ Phys.\ J.\  C {\bf 22} (2001) 379
  [arXiv:hep-th/9803001].
  
\bibitem{sonwilson}
  J.~Sonnenschein,  
\emph{``What does the string/gauge correspondence teach us about Wilson loops?''}
  [arXiv:hep-th/0003032]
  
\bibitem{Hartnoll:2009sz}
  S.~A.~Hartnoll,
\emph{``Lectures on holographic methods
for condensed matter physics,''}
	Class.\ quant.\ Grav.{\bf 26} :224002 (2009)
  [arXiv:0903.3246 [hep-th]].
  
\bibitem{Son:2005rv}
  D.~T.~Son and M.~Wingate,
\emph{``General coordinate invariance
and conformal invariance in
nonrelativistic physics: Unitary Fermi
gas,''}
  Annals Phys.\  {\bf 321}, 197 (2006)
  [arXiv:cond-mat/0509786].

\bibitem{Nishida:2007pj}
  Y.~Nishida and D.~T.~Son,
\emph{``Nonrelativistic conformal field
theories,''}
  Phys.\ Rev.\  D {\bf 76}, 086004 (2007)
  [arXiv:0706.3746 [hep-th]].

\bibitem{Son:2008ye}
  D.~T.~Son,
  \emph{``Toward an AdS/cold atoms
  correspondence: a geometric realization of the
  Schroedinger symmetry,''}
  Phys.\ Rev.\  D {\bf 78}, 046003 (2008)
  [arXiv:0804.3972 [hep-th]].

\bibitem{Balasubramanian:2008dm}
  K.~Balasubramanian and J.~McGreevy,
	\emph{``Gravity duals for
non-relativistic CFTs,''}
  Phys.\ Rev.\ Lett.\  {\bf 101}, 061601 (2008)
  [arXiv:0804.4053 [hep-th]].

\bibitem{Akhavan:2008ep}
  A.~Akhavan, M.~Alishahiha, A.~Davody and A.~Vahedi,
	\emph{``Non-relativistic CFT and
Semi-classical Strings,''}
  JHEP {\bf 0903} (2009) 053
  [arXiv:0811.3067 [hep-th]].

\bibitem{Kluson:2010prd}
	J. ~Kluso\v{n},
	\emph{``Open String in Non-Relativistic Background,''}
	Phys.\ Rev. {\bf D81}, 106006 (2010) 
	[arXiv:0912.4587 [hep-th]].

\bibitem{Balabsnumber2010}
	K. ~Balasubramanian and J. ~McGreevy,
	\emph{``The particle number in Galilean holography,''}
	JHEP {\bf 1101} (2011) 137
	[arXiv:1007.2184 [hep-th]].

\bibitem{bachas}
  C.~Bachas,
  \emph{``Convexity Of The Quarkonium Potential,''}
  Phys.\ Rev.\  D {\bf 33} (1986) 2723.
  
\bibitem{nunez}
	C. ~Nunez, M. ~Piai, A. ~Rago,
	\emph{``Wilson Loops in string duals of Walking and Flavored Systems,''}
	Phys.\ Rev.\ D {\bf 81}, 086001 (2010)
	[arXiv:0909.0748 [hep-th]].

\bibitem{filho}
	H. ~Boschi-Filho and N. R. F. ~Braga,
	\emph{``Wilson Loops for a quark anti-quark pair in D3-brane
space,''}
	\textit{JHEP} {\bf 03} (2005) 051 
	[arXiv:hep-th/0411135].
	
\bibitem{caceres}
	E. ~C\'{a}ceres, M. ~Natsuume and T. ~Okamura, 
	\emph{``Screening length in plasma winds,''}
	\textit{JHEP} \textbf{0610} (2006) 011 
	[arXiv:hep-th/0607233].

\bibitem{kinar}
	Y. ~Kinar, E. ~Schreiber, J. ~Sonnenschein,
	\emph{``$Q \bar{Q}$ Potential from Strings in Curved Spacetime - Classical Results,''}
	Nucl.\ Phys.\ {\bf B} 566 (2000) 103-125
	[arXiv:hep-th/9811192].

\bibitem{arias}
	R. E. ~Arias and G. A. ~Silva,
	\emph{``Wilson loops stability in the gauge/string correspondence,''}
	\textit{JHEP} \textbf{1001} (2010)023
	[arXiv:0911.0662 [hep-th]].

\bibitem{yoshida-hartnoll}
	S. A. ~Hartnoll and K. ~Yoshida,
	\emph{``Families of IIB duals for nonrelativistic CFTs,''}
	\textit{JHEP} \textbf{0812} (2008)071
	[arXiv:0810.0298 [hep-th]]
	
\end{thebibliography}
\end{document}